\documentclass[prd,aps,showpacs,epsf,floats,onecolumn]{revtex4}%
\usepackage{amssymb}
\usepackage{amsfonts}
\usepackage{amsmath}
\usepackage{graphicx}%
\providecommand{\U}[1]{\protect\rule{.1in}{.1in}}
\begin{document}
\title{\textbf{Information Geometric Measures of Complexity with Applications to
Classical and Quantum Physical Settings }}
\author{\textbf{Carlo Cafaro}$^{1}$\textbf{ and Sean A. Ali}$^{2}$}
\affiliation{$^{1}$SUNY Polytechnic Institute, 12203 Albany, New York, USA}
\affiliation{$^{2}$Albany College of Pharmacy and Health Sciences, 12208 Albany, New York, USA}

\begin{abstract}
We discuss the fundamental theoretical framework together with numerous
results obtained by the authors and colleagues over an extended period of
investigation on the Information Geometric Approach to Chaos (IGAC).

\end{abstract}

\pacs{Chaos (05.45.-a), Complexity (89.70.Eg), Entropy (89.70.Cf), Inference Methods
(02.50.Tt), Information Theory (89.70.+c), Probability Theory (02.50.Cw),
Riemannian Geometry (02.40.Ky).}
\maketitle

\section{Theoretical background}

Statistical models are employed to formulate probabilistic descriptions of
systems of arbitrary nature when only partial knowledge about the system is
available. Indeed, in recent years, methods of entropic inference
\cite{caticha12} have been utilized in conjunction with information geometry
(IG) \cite{amari} for the purpose of developing complexity indicators of
statistical models. From the perspective of this hybrid framework, such
complexity indicators can be understood as being quantitative measures that
describe the complication of inferring macroscopic predictions about
statistical models. In this context, the difficulty of making macroscopic
predictions is attributed to the fact that statistical models intrinsically
reflect only partial information about the microscopic degrees of freedom of
the system being modeled. Initial theoretical investigation in this direction,
quoted as the \emph{Information Geometric Approach to Chaos }(IGAC), was
originally proposed by Cafaro in his physics Ph.D. doctoral dissertation in
Ref. \cite{cafarophd}.

A general summary of the IGAC\ framework is described as follows
\cite{ali18,felice18}: upon identifying the microscopic degrees of freedom of
a complex system, one must obtain data and choose important information
constraints on the system. Entropic methods are then utilized to obtain an
initial, static statistical model of the system. In this way, the system is
described by a statistical model specified in terms of probability
distributions that are characterized by statistical macrovariables. The
statistical macrovariables are determined by the data and the specific
functional expression of the information constraints used to implement
statistical inferences. The next step in the theoretical scheme is concerned
with the temporal evolution of the system. If it is assumed that the system
changes, then the corresponding statistical model evolves from its initial to
final configurations in a manner specified by Entropic Dynamics (ED,
\cite{catichaED}). The ED framework can be viewed as a form of constrained
information dynamics that is formulated on statistical manifolds, the elements
of which are probability distributions. These distributions in turn are in
one-to-one relation with an appropriate set of statistical macrovariables that
determine a parameter space, where the latter serves to provide a suitable
parametrization of points on the original statistical manifold.

Within the context of ED, the change of probability distributions is described
in terms of a principle of entropic inference. Specifically, beginning with a
known initial configuration, change toward the final configuration happens by
the maximization of the logarithmic relative entropy (known as the Maximum
relative Entropy method - or MrE method in brief, \cite{caticha12}) between
any two successive intermediate configurations. We emphasize that
ED\ specifies the \emph{expected} rather than the \emph{actual} dynamical
paths of the system. Inferences within the ED framework depends on the data
and functional form of the selected information constraints used in the MrE
algorithm. Indeed, modeling strategies of this kind can only be corroborated
\emph{a posteriori}. This fact implies that in the event inferred predictions
fail to match experimental measurements, a new set of information constraints
should be chosen. This feature of the MrE algorithm is of critical
significance and was recently re-examined by Cafaro and Ali in Ref.
\cite{cafaropre16}.

The change of probability distributions characterized by the maximization
algorithm outlined above prescribes a geodesic evolution for the statistical
macrovariables \cite{caticha12}. The Fisher-Rao information metric
\cite{amari} yields a measure of distance between any two dissimilar
probability distributions on a statistical manifold. The notion of distance
between elements of a statistical manifold can be regarded as the degree of
distinguishability between any two different probability distribution
functions. Once the information metric has be obtained, differential geometric
techniques can be readily applied to study the geometry of the curved
statistical manifold. Broadly speaking, standard Riemannian geometric
quantities such as Christoffel connection coefficients of the second kind,
Riemannian curvature tensor, Ricci tensor, Ricci scalar curvature, Weyl
anisotropy tensor, sectional curvatures, Killing fields and Jacobi fields
(including the IG\ analogue of Lyapunov exponents) can be calculated in the
usual fashion. In particular, the \emph{chaoticity} (i.e.\textbf{,}
\emph{temporal complexity}) of such statistical models can be\textbf{
}analyzed via appropriately selected indicators, such as the signs of the
Ricci scalar and sectional curvatures of the statistical manifold,
non-vanishing Weyl anisotropy tensor, the asymptotic temporal behavior of
Jacobi fields and the existence of Killing vectors. Along with the various
indicators mentioned above, the notion of complexity within the
IGAC\ framework can also be characterized by the \emph{Information Geometric
Entropy} (IGE), originally proposed in \cite{cafarophd}. We make reference to
the Ali-Cafaro effort in Ref. \cite{ali17} for a more extensive summary of the
IGAC framework that incorporates a set of remarks on entropic evolution and
the MrE algorithm. For a presentation of alternative information geometric
descriptions of complexity, we suggest the investigation by Felice, Cafaro and
Mancini in Ref. \cite{felice18}. While we certainly appreciate the power of
the synthetic, non-component approach to tensors analysis commonly used in
theoretical physics (for instance, see Refs. \cite{cafaro07A}), we have
nevertheless chosen to employ the component approach in the present paper.\ In
our opinion, the applied nature of our works can be formulated and analyzed
more efficiently (and transparently) within the component approach to tensor calculus.

In the next section, we introduce suitable indicators of complexity within the IGAC.

\section{Indicators of complexity}

In this section, we introduce three indicators of complexity within the
context of IGAC framework. We present the IGE, the curvature of the
statistical manifold and finally, the notion of Jacobi fields arising from the
equation of geodesic deviation.

\subsection{Information geometric entropy}

We begin this subsection by discussing the IGE. Assuming the elements
$\left\{  p\left(  x;\theta\right)  \right\}  $ of an $n$-dimensional
statistical manifold $\mathcal{M}_{s}$ are parametrized by $n$, \emph{real}
valued variables $\left(  \theta^{1}\text{,..., }\theta^{n}\right)  $, the
statistical manifold is defined by the set,
\begin{equation}
\mathcal{M}_{s}\overset{\text{def}}{=}\left\{  p\left(  x;\theta\right)
:\theta=\left(  \theta^{1}\text{,..., }\theta^{n}\right)  \in\mathcal{D}%
_{\boldsymbol{\theta}}^{\left(  \text{tot}\right)  }\right\}  \text{.}
\label{smanifold}%
\end{equation}
We point out that the quantity $x$ in Eq. (\ref{smanifold}) are microvariables
belonging to the microspace $\mathcal{X}$, whereas the macrovariables $\theta$
appearing in Eq. (\ref{smanifold}) are elements of the parameter space
$\mathcal{D}_{\boldsymbol{\theta}}^{\left(  \text{tot}\right)  }$ defined by,%
\begin{equation}
\mathcal{D}_{\boldsymbol{\theta}}^{\left(  \text{tot}\right)  }\overset
{\text{def}}{=}{\bigotimes\limits_{j=1}^{n}}\mathcal{I}_{\theta^{j}}=\left(
\mathcal{I}_{\theta^{1}}\otimes\mathcal{I}_{\theta^{2}}\text{...}%
\otimes\mathcal{I}_{\theta^{n}}\right)  \subseteq\mathbb{R}^{n}\text{.}
\label{dtot}%
\end{equation}
The quantity $\mathcal{I}_{\theta^{j}}$ with $1\leq j\leq n$ in Eq.
(\ref{dtot}) is a subset of $\mathbb{R}^{n}$ that denotes the full range of
admissible values of the statistical macrovariables $\theta^{j}$. The IGE
serves as an indicator of temporal complexity associated with geodesic paths
in the IGAC framework. The IGE\ is given by,
\begin{equation}
\mathcal{S}_{\mathcal{M}_{s}}\left(  \tau\right)  \overset{\text{def}}{=}%
\log\widetilde{vol}\left[  \mathcal{D}_{\boldsymbol{\theta}}\left(
\tau\right)  \right]  \text{,} \label{IGE}%
\end{equation}
where the average dynamical statistical volume\textbf{\ }$\widetilde
{vol}\left[  \mathcal{D}_{\boldsymbol{\theta}}\left(  \tau\right)  \right]  $
is defined as,
\begin{equation}
\widetilde{vol}\left[  \mathcal{D}_{\boldsymbol{\theta}}\left(  \tau\right)
\right]  \overset{\text{def}}{=}\frac{1}{\tau}\int_{0}^{\tau}vol\left[
\mathcal{D}_{\boldsymbol{\theta}}\left(  \tau^{\prime}\right)  \right]
d\tau^{\prime}\text{.} \label{rhs}%
\end{equation}
We remark that the temporal averaging operation is denoted by the tilde symbol
in Eq. (\ref{rhs}). Furthermore, the volume\textbf{\ }$vol\left[
\mathcal{D}_{\boldsymbol{\theta}}\left(  \tau^{\prime}\right)  \right]
$\ appearing on the RHS of Eq. (\ref{rhs}) is given by,
\begin{equation}
vol\left[  \mathcal{D}_{\boldsymbol{\theta}}\left(  \tau^{\prime}\right)
\right]  \overset{\text{def}}{=}\int_{\mathcal{D}_{\boldsymbol{\theta}}\left(
\tau^{\prime}\right)  }\rho\left(  \theta^{1}\text{,..., }\theta^{n}\right)
d^{n}\theta\text{,} \label{v}%
\end{equation}
where $\rho\left(  \theta^{1}\text{,..., }\theta^{n}\right)  $ is known as the
Fisher density and is equal to the square root of the determinant $g\left(
\theta\right)  $ of the Fisher-Rao information metric tensor $g_{\mu\nu
}\left(  \theta\right)  $,
\begin{equation}
\rho\left(  \theta^{1}\text{,..., }\theta^{n}\right)  \overset{\text{def}}%
{=}\sqrt{g\left(  \theta\right)  }\text{.}%
\end{equation}
The Fisher-Rao information metric tensor $g_{\mu\nu}\left(  \theta\right)  $
is defined as,%
\begin{equation}
g_{\mu\nu}\left(  \theta\right)  \overset{\text{def}}{=}\int p\left(
x|\theta\right)  \partial_{\mu}\log p\left(  x|\theta\right)  \partial_{\nu
}\log p\left(  x|\theta\right)  dx\text{,} \label{FRyou}%
\end{equation}
where $\mu$, $\nu=1$,..., $n$ for an $n$-dimensional manifold and
$\partial_{\mu}\overset{\text{def}}{=}\frac{\partial}{\partial\theta^{\mu}}$.
The volume $vol\left[  \mathcal{D}_{\boldsymbol{\theta}}\left(  \tau^{\prime
}\right)  \right]  $ in Eq. (\ref{v}) can be recast in a more crystalline
manner for cases involving statistical manifolds whose information metric
tensor has a determinant that can be expressed in a factorized form as
follows,
\begin{equation}
g\left(  \theta\right)  =g\left(  \theta^{1}\text{,..., }\theta^{n}\right)
={\prod\limits_{j=1}^{n}}g_{j}\left(  \theta^{j}\right)  \text{.}%
\end{equation}
By using the factorized form of the determinant, the IGE appearing in Eq.
(\ref{IGE}) can be expressed as
\begin{equation}
\mathcal{S}_{\mathcal{M}_{s}}\left(  \tau\right)  =\log\left\{  \frac{1}{\tau
}\int_{0}^{\tau}\left[  {\prod\limits_{j=1}^{n}}\left(  \int_{\tau_{0}}%
^{\tau_{0}+\tau^{\prime}}\sqrt{g_{j}\left[  \theta^{j}\left(  \xi\right)
\right]  }\frac{d\theta^{j}}{d\xi}d\xi\right)  \right]  d\tau^{\prime
}\right\}  \text{.} \label{IGEmod}%
\end{equation}
Within the IGAC framework, the leading asymptotic behavior of $\mathcal{S}%
_{\mathcal{M}_{s}}\left(  \tau\right)  $ in\ Eq. (\ref{IGEmod}) is employed to
specify the complexity of the statistical model under investigation.
Therefore, it is quite instructive to take into consideration \ the quantity
\begin{equation}
\mathcal{S}_{\mathcal{M}_{s}}^{\left(  \text{asymptotic}\right)  }\left(
\tau\right)  \approx\lim_{\tau\rightarrow\infty}\left[  \mathcal{S}%
_{\mathcal{M}_{s}}\left(  \tau\right)  \right]  \text{,}%
\end{equation}
that is, the leading asymptotic term in the expression of the IGE. The
integration space $\mathcal{D}_{\theta}\left(  \tau^{\prime}\right)  $ in Eq.
(\ref{v}) is defined by
\begin{equation}
\mathcal{D}_{\boldsymbol{\theta}}\left(  \tau^{\prime}\right)  \overset
{\text{def}}{=}\left\{  \theta:\theta^{j}\left(  \tau_{0}\right)  \leq
\theta^{j}\leq\theta^{j}\left(  \tau_{0}+\tau^{\prime}\right)  \right\}
\text{,} \label{is}%
\end{equation}
where $\theta^{j}=\theta^{j}\left(  \xi\right)  $ with $\tau_{0}\leq\xi
\leq\tau_{0}+\tau^{\prime}$ and $\tau_{0}$ denotes the initial value of the
affine parameter $\xi$ such that,
\begin{equation}
\frac{d^{2}\theta^{j}\left(  \xi\right)  }{d\xi^{2}}+\Gamma_{ik}^{j}%
\frac{d\theta^{i}}{d\xi}\frac{d\theta^{k}}{d\xi}=0\text{.} \label{ge}%
\end{equation}
The domain of integration $\mathcal{D}_{\boldsymbol{\theta}}\left(
\tau^{\prime}\right)  $ in Eq. (\ref{is}) is an $n$-dimensional subspace of
$\mathcal{D}_{\boldsymbol{\theta}}^{\left(  \text{tot}\right)  }$. The
elements of $\mathcal{D}_{\boldsymbol{\theta}}^{\left(  \text{tot}\right)  }$
are $n$-dimensional macrovariables $\left\{  \theta\right\}  $ whose
components $\theta^{j}$ are bounded by the limits of integration $\theta
^{j}\left(  \tau_{0}\right)  $ and $\theta^{j}\left(  \tau_{0}+\tau^{\prime
}\right)  $. The temporal functional form of such limits is determined by the
integration of the geodesic equations in Eq. (\ref{ge}).

The IGE evaluated at a particular instant is specified by the logarithm of the
volume of the effective parameter space probed by the system at that instant.
In order to coarse-grain the possibly very complex details of the entropic
dynamical characterization of the system however, the process of temporal
averaging has been employed. Moreover, in order to remove the effects of
potential transient features that may enter the calculation of the expected
value of the volume of the effective parameter space, only its asymptotic
temporal behavior is taken into consideration. For these reasons, it is
evident that the IGE serves as an asymptotic, coarse-grained complexity
indicator of dynamical systems in the presence of partial information. For
additional specifics concerning the IGE, we refer the interested reader to
Refs. \cite{cafaroamc10,ali17}.

We emphasize that it would be interesting to characterize the tendency to
increase of the entropy of a physical system that approaches equilibrium as
specified by the Boltzmann $H$ theorem and the second law of thermodynamics
\cite{kittel} from an information geometric perspective. For a recent
information geometric interpretation of the entropy production, we refer to
Ref. \cite{ito20}. In particular, to understand the possible link between the
IGE and the Boltzmann-Shannon entropy, it would be important to study the
Kaniadakis\textbf{ }$\mathcal{S}_{\kappa}$\textbf{ }entropy (with $\kappa$
being the so-called deformation parameter) and comprehend how the statistical
mechanics based on\textbf{ }$\mathcal{S}_{\kappa}$\textbf{ }can be regarded as
a natural generalization of the equilibrium Boltzmann-Gibbs statistical
mechanics \cite{kaniadakis02}. We leave these intriguing lines of
investigations to future efforts.

\subsection{Curvature}

We present here the notion of curvature of statistical manifolds. We begin by
recalling that an $n$-dimensional, $%
\mathbb{C}
^{\infty}$ differentiable manifold is defined by a set of points $\mathcal{M}$
endowed with coordinate systems $\mathcal{C}_{\mathcal{M}}$ fulfilling the
following two requirements: 1) each element $c\in\mathcal{C}_{\mathcal{M}}$ is
a one-to-one mapping from $\mathcal{M}$ to an open subset of $%
\mathbb{R}
^{n}$; 2) given any one-to-one mapping $\eta:\mathcal{M}\rightarrow%
\mathbb{R}
^{n}$, we have that $\forall c\in\mathcal{C}_{\mathcal{M}}$, $\eta
\in\mathcal{C}_{\mathcal{M}}\Leftrightarrow\eta\circ c^{-1}$ is a $%
\mathbb{C}
^{\infty}$ diffeomorphism.

In this paper, we focus on Riemannian manifolds $\left(  \mathcal{M}\text{,
}g\right)  $ where the points of $\mathcal{M}$ are probability distribution
functions. It is\textbf{ }worth noting that the manifold structure of
$\mathcal{M}$ is insufficient to specify in a unique manner the Riemannian
metric $g$. On a formal level, an infinite number of Riemannian metrics can be
defined on the manifold $\mathcal{M}$ . In the context of information geometry
however, the selection of the Fisher-Rao information metric (see Eq.
(\ref{FRyou})) as the metric underlying the Riemannian geometry of probability
distributions \cite{amari,fisher,rao} serves as a primary working assumption.
The characterization theorem attributed to Cencov \cite{cencov} gives
significant support for this particular choice of metric. In this
characterization theorem, Cencov demonstrates that, up to any arbitrary
constant scale factor, the information metric is the only Riemannian metric
that is invariant under congruent embeddings (that is, under a family of
probabilistically meaningful\textbf{ }mappings) of Markov morphism
\cite{cencov, campbell}.

Upon introducing the Fisher-Rao information metric $g_{\mu\nu}\left(
\theta\right)  $ in Eq. (\ref{FRyou}), standard differential geometric
techniques can be used on\textbf{ }the space of probability distributions to
describe the geometry of the statistical manifold $\mathcal{M}_{s}$. The Ricci
scalar curvature\textbf{ }$\mathcal{R}_{\mathcal{M}_{s}}$\textbf{ }is one
example of such a geometric property, where $\mathcal{R}_{\mathcal{M}_{s}}$ is
defined as \cite{weinberg},%
\begin{equation}
\mathcal{R}_{\mathcal{M}_{s}}\overset{\text{def}}{=}g^{\mu\nu}\mathcal{R}%
_{\mu\nu}\text{,} \label{ricci-scalar}%
\end{equation}
where $g^{\mu\nu}g_{\nu\rho}=\delta_{\rho}^{\mu}$ and $g^{\mu\nu}=\left(
g_{\mu\nu}\right)  ^{-1}$. The Ricci tensor $\mathcal{R}_{\mu\nu}$
appearing\textbf{ }in Eq. (\ref{ricci-scalar}) is given as \cite{weinberg},%
\begin{equation}
\mathcal{R}_{\mu\nu}\overset{\text{def}}{=}\partial_{\gamma}\Gamma_{\mu\nu
}^{\gamma}-\partial_{\nu}\Gamma_{\mu\lambda}^{\lambda}+\Gamma_{\mu\nu}%
^{\gamma}\Gamma_{\gamma\eta}^{\eta}-\Gamma_{\mu\gamma}^{\eta}\Gamma_{\nu\eta
}^{\gamma}\text{.} \label{ricci-tensor}%
\end{equation}
The Christoffel connection coefficients $\Gamma_{\mu\nu}^{\rho}$ of the second
kind that specify the Ricci tensor in Eq. (\ref{ricci-tensor}) are
\cite{weinberg},
\begin{equation}
\Gamma_{\mu\nu}^{\rho}\overset{\text{def}}{=}\frac{1}{2}g^{\rho\sigma}\left(
\partial_{\mu}g_{\sigma\nu}+\partial_{\nu}g_{\mu\sigma}-\partial_{\sigma
}g_{\mu\nu}\right)  \text{.} \label{connection}%
\end{equation}
Next we consider geodesic curves on statistical manifolds\textbf{.} A geodesic
on an $n$-dimensional statistical manifold $\mathcal{M}_{s}$ can be
interpreted as the maximum probability trajectory explored by a complex system
during its change from an initial $\theta_{\text{initial}}$ to final
macrostates $\theta_{\text{final}}$, respectively. Each point along a geodesic
path corresponds to a macrostate specified by the macroscopic\textbf{
}variables $\theta=\left(  \theta^{1}\text{,..., }\theta^{n}\right)  $. In the
context of ED, each component $\theta^{j}$ with $j=1$,..., $n$ is a solution
of the geodesic equation \cite{catichaED},%
\begin{equation}
\frac{d^{2}\theta^{k}}{d\xi^{2}}+\Gamma_{lm}^{k}\frac{d\theta^{l}}{d\xi}%
\frac{d\theta^{m}}{d\xi}=0\text{.}%
\end{equation}
At this juncture, we reiterate the fact that each macrostate $\theta$ is in
one-to-one correspondence with the probability distribution $p\left(
x|\theta\right)  $, with the latter characterizing a distribution of the
microstates $x$. It is useful to recognize that the scalar curvature
$\mathcal{R}_{\mathcal{M}_{s}}$ can be readily recast as the sum of sectional
curvatures $\mathcal{K}\left(  e_{\rho}\text{, }e_{\sigma}\right)  $ of all
tangent space planes $T_{p}\mathcal{M}_{s}$ with $p\in\mathcal{M}_{s}$ spanned
by pairs of orthonormal basis vectors $\left\{  e_{\rho}=\partial
_{\theta_{\rho}(p)}\right\}  $,%
\begin{equation}
\mathcal{R}_{\mathcal{M}_{s}}\overset{\text{def}}{=}\mathcal{R}_{\text{
}\alpha}^{\alpha}\overset{\text{def}}{=}\sum_{\rho\neq\sigma}\mathcal{K}%
\left(  e_{\rho}\text{, }e_{\sigma}\right)  \text{,} \label{Ricci}%
\end{equation}
where $\mathcal{K}\left(  a\text{, }b\right)  $ is given by \cite{weinberg},%
\begin{equation}
\mathcal{K}\left(  a\text{, }b\right)  \overset{\text{def}}{=}\frac
{\mathcal{R}_{\mu\nu\rho\sigma}a^{\mu}b^{\nu}a^{\rho}b^{\sigma}}{\left(
g_{\mu\sigma}g_{\nu\rho}-g_{\mu\rho}g_{\nu\sigma}\right)  a^{\mu}b^{\nu
}a^{\rho}b^{\sigma}}\text{,} \label{sectionK}%
\end{equation}
with,%
\begin{equation}
a\overset{\text{def}}{=}\sum_{\rho}\left\langle a\text{, }e^{\rho
}\right\rangle e_{\rho}\text{, }b\overset{\text{def}}{=}\sum_{\rho
}\left\langle b\text{, }e^{\rho}\right\rangle e_{\rho}\text{, and
}\left\langle e_{\rho}\text{, }e^{\sigma}\right\rangle \overset{\text{def}}%
{=}\delta_{\rho}^{\sigma}\text{.}%
\end{equation}
We observe that the Riemann curvature tensor $\mathcal{R}_{\alpha\beta
\rho\sigma}$ \cite{weinberg} is fully determined by the sectional curvatures
$\mathcal{K}\left(  e_{\rho}\text{, }e_{\sigma}\right)  $ where,%
\begin{equation}
\mathcal{R}^{\alpha}\,_{\beta\rho\sigma}\overset{\text{def}}{=}g^{\alpha
\gamma}\mathcal{R}_{\gamma\beta\rho\sigma}\overset{\text{def}}{=}%
\partial_{\sigma}\Gamma_{\text{ \ }\beta\rho}^{\alpha}-\partial_{\rho}%
\Gamma_{\text{ \ }\beta\sigma}^{\alpha}+\Gamma^{\alpha}\,_{\lambda\sigma
}\Gamma^{\lambda}\,_{\beta\rho}-\Gamma^{\alpha}\,_{\lambda\rho}\Gamma
^{\lambda}\,_{\beta\sigma}\text{.}%
\end{equation}
The negativity of the Ricci scalar\textbf{ }curvature $\mathcal{R}%
_{\mathcal{M}_{s}}$ is a strong (i.e., a sufficient but not necessary)
criterion of local\textbf{ }dynamical instability. Moreover, the compactness
of the manifold $\mathcal{M}_{s}$\ is required to specify genuine chaotic
(that is, temporally complex) dynamical systems. In particular, it is evident
from Eq. (\ref{Ricci}) that the negativity of $\mathcal{R}_{\mathcal{M}_{s}}$
imply that negative principal curvatures (i.e.\textbf{,} extrema of sectional
curvatures) are more dominant than positive ones. For this reason\textbf{,}
the negativity of $\mathcal{R}_{\mathcal{M}_{s}}$ is a sufficient but not
necessary requirement for local instability of geodesic flows on statistical
manifolds. It is worth mentioning the possible circumstance of scenarios in
which negative sectional curvatures are present, but the positive curvatures
dominate in the sum of Eq. (\ref{Ricci}) such that $\mathcal{R}_{\mathcal{M}%
_{s}}$ is a non-negative quantity despite flow instability in those
directions. For additional mathematical considerations related to the concept
of curvature in differential geometry, we suggest Ref. \cite{lee}.

\subsection{Jacobi fields}

We introduce here the concept of the Jacobi vector field. It is worth noting
that the analysis of stability/instability arising in natural (geodesic)
evolutions is readily accomplished by means of the Jacobi-Levi-Civita (JLC)
equation for geodesic deviation. This equation is familiar in both theoretical
physics (for example, in the case of General Relativity) as well as in
Riemannian geometry. The JLC equation describes in a covariant manner, the
degree to which neighboring geodesics locally scatter. In particular, the JLC
equation effectively connects the curvature properties of an underlying
manifold to the stability/instability of the geodesic flow induced thereupon.
Indeed, the JLC equation provides a window into a diverse and mostly
unexplored field of study concerning the connections among topology, geometry
and geodesic instability, and thus to complexity and chaoticity. The use the
JLC equation in the setting of information geometry originally appeared in
Ref. \cite{cafaropd07}.

In what follows we take into consideration two neighboring geodesic paths
$\theta^{\alpha}\left(  \xi\right)  $ and $\theta^{\alpha}\left(  \xi\right)
+\delta\theta^{\alpha}\left(  \xi\right)  $, where the quantity $\xi$\textbf{
}denotes an affine parameter satisfying the geodesic equations,%
\begin{equation}
\frac{d^{2}\theta^{\alpha}}{d\xi^{2}}+\Gamma_{\beta\gamma}^{\alpha}\left(
\theta\right)  \frac{d\theta^{\beta}}{d\xi}\frac{d\theta^{\gamma}}{d\xi
}=0\text{,}\label{ge1}%
\end{equation}
and%
\begin{equation}
\frac{d^{2}\left[  \theta^{\alpha}+\delta\theta^{\alpha}\right]  }{d\xi^{2}%
}+\Gamma_{\beta\gamma}^{\alpha}\left(  \theta+\delta\theta\right)
\frac{d\left[  \theta^{\beta}+\delta\theta^{\beta}\right]  }{d\xi}%
\frac{d\left[  \theta^{\gamma}+\delta\theta^{\gamma}\right]  }{d\xi}%
=0\text{,}\label{ge2}%
\end{equation}
respectively. Noting that to first order in $\delta\theta^{\alpha}$,
\begin{equation}
\Gamma_{\beta\gamma}^{\alpha}\left(  \theta+\delta\theta\right)  \approx
\Gamma_{\beta\gamma}^{\alpha}\left(  \theta\right)  +\partial_{\eta}%
\Gamma_{\beta\gamma}^{\alpha}\delta\theta^{\eta}\text{,}%
\end{equation}
after some algebraic calculations, to first order in $\delta\theta^{\alpha}$
Eq. (\ref{ge2}) becomes%
\begin{equation}
\frac{d^{2}\theta^{\alpha}}{d\xi^{2}}+\frac{d^{2}\left(  \delta\theta^{\alpha
}\right)  }{d\xi^{2}}+\Gamma_{\beta\gamma}^{\alpha}\left(  \theta\right)
\frac{d\theta^{\beta}}{d\xi}\frac{d\theta^{\gamma}}{d\xi}+2\Gamma_{\beta
\gamma}^{\alpha}\left(  \theta\right)  \frac{d\theta^{\beta}}{d\xi}%
\frac{d\left(  \delta\theta^{\gamma}\right)  }{d\xi}+\partial_{\eta}%
\Gamma_{\beta\gamma}^{\alpha}\left(  \theta\right)  \delta\theta^{\eta}%
\frac{d\theta^{\beta}}{d\xi}\frac{d\theta^{\gamma}}{d\xi}=0\text{.}\label{ge3}%
\end{equation}
The equation of geodesic deviation can be found by subtracting Eq. (\ref{ge1})
from Eq. (\ref{ge3}),%
\begin{equation}
\frac{d^{2}\left(  \delta\theta^{\alpha}\right)  }{d\xi^{2}}+2\Gamma
_{\beta\gamma}^{\alpha}\left(  \theta\right)  \frac{d\theta^{\beta}}{d\xi
}\frac{d\left(  \delta\theta^{\gamma}\right)  }{d\xi}+\partial_{\eta}%
\Gamma_{\beta\gamma}^{\alpha}\left(  \theta\right)  \delta\theta^{\eta}%
\frac{d\theta^{\beta}}{d\xi}\frac{d\theta^{\gamma}}{d\xi}=0\text{.}\label{ge4}%
\end{equation}
Equation (\ref{ge4}) can be conveniently recast via the covariant derivatives
(see Ref. \cite{ohanian}, for instance) along the curve $\theta^{\alpha
}\left(  \xi\right)  $,%
\begin{align}
\frac{D^{2}\left(  \delta\theta^{\alpha}\right)  }{D\xi^{2}} &  =\frac
{d^{2}\left(  \delta\theta^{\alpha}\right)  }{d\xi^{2}}+\partial_{\beta}%
\Gamma_{\rho\sigma}^{\alpha}\frac{d\theta^{\beta}}{d\xi}\delta\theta^{\rho
}\frac{d\theta^{\sigma}}{d\xi}+2\Gamma_{\rho\sigma}^{\alpha}\frac{d\left(
\delta\theta^{\rho}\right)  }{d\xi}\frac{d\theta^{\sigma}}{d\xi}+\nonumber\\
& \nonumber\\
&  -\Gamma_{\rho\sigma}^{\alpha}\Gamma_{\kappa\lambda}^{\sigma}\delta
\theta^{\rho}\frac{d\theta^{\kappa}}{d\xi}\frac{d\theta^{\lambda}}{d\xi
}+\Gamma_{\rho\sigma}^{\alpha}\Gamma_{\kappa\lambda}^{\rho}\delta
\theta^{\kappa}\frac{d\theta^{\lambda}}{d\xi}\frac{d\theta^{\sigma}}{d\xi
}\text{.}\label{ge5}%
\end{align}
The covariant derivative is defined as $D_{\xi}\delta\theta^{\alpha}%
\overset{\text{def}}{=}\partial_{\xi}\delta\theta^{\alpha}+\Gamma_{\xi\kappa
}^{\alpha}\delta\theta^{\kappa}$ with $D_{\xi}\overset{\text{def}}{=}D/D\xi$
and $\partial_{\xi}\overset{\text{def}}{=}\partial/\partial\xi$, respectively.
By combining Eqs. (\ref{ge4}) and (\ref{ge5}), and performing some tensor
algebra calculations, we get%
\begin{equation}
\frac{D^{2}\left(  \delta\theta^{\alpha}\right)  }{D\xi^{2}}=\left(
\partial_{\rho}\Gamma_{\eta\sigma}^{\alpha}-\partial_{\eta}\Gamma_{\rho\sigma
}^{\alpha}+\Gamma_{\lambda\sigma}^{\alpha}\Gamma_{\eta\rho}^{\lambda}%
-\Gamma_{\eta\lambda}^{\alpha}\Gamma_{\rho\sigma}^{\lambda}\right)
\delta\theta^{\eta}\frac{d\theta^{\rho}}{d\xi}\frac{d\theta^{\sigma}}{d\xi
}\text{.}\label{ge6}%
\end{equation}
Finally, the geodesic deviation equation expressed in component form becomes%
\begin{equation}
\frac{D^{2}J^{\alpha}}{D\xi^{2}}+\mathcal{R}_{\rho\eta\sigma}^{\alpha}%
\frac{d\theta^{\rho}}{d\xi}J^{\eta}\frac{d\theta^{\sigma}}{d\xi}%
=0\text{,}\label{JLC}%
\end{equation}
where $J^{\alpha}\overset{\text{def}}{=}\delta\theta^{\alpha}$ is the $\alpha
$-component of the Jacobi vector field \cite{weinberg}. Equation (\ref{JLC})
is known formally as the JLC equation.\textbf{ }We observe from the JLC
equation in Eq. (\ref{JLC}) that neighboring geodesics accelerate relative to
each other at a rate measured in a direct manner by the Riemannian curvature
tensor $R_{\alpha\beta\gamma\delta}$. The quantity $J^{\alpha}$ is defined as,%
\begin{equation}
J^{\alpha}=\delta\theta^{\alpha}\overset{\text{def}}{=}\delta_{\phi}%
\theta^{\alpha}=\left(  \frac{\partial\theta^{\alpha}\left(  \xi\text{; }%
\phi\right)  }{\partial\phi}\right)  _{\tau=\text{constant}}\delta\phi
\text{,}\label{jacobi}%
\end{equation}
where $\left\{  \theta^{\mu}\left(  \xi\text{; }\phi\right)  \right\}  $
denotes the one-parameter $\phi$ family of geodesics whose evolution is
specified by means of the affine parameter $\xi$. The Jacobi vector field
intensity $J_{\mathcal{M}_{s}}$ on the manifold $\mathcal{M}_{s}$ is given by,%
\begin{equation}
J_{\mathcal{M}_{s}}\overset{\text{def}}{=}\left(  J^{\alpha}g_{\alpha\beta
}J^{\beta}\right)  ^{1/2}\text{.}\label{Jintensity}%
\end{equation}
In general, the JLC equation is intractable even at low dimensions. However,
in the case of isotropic manifolds, it reduces to
\begin{equation}
\frac{D^{2}J^{\mu}}{D\xi^{2}}+\mathcal{K}J^{\mu}=0\text{.}%
\label{geo-deviation}%
\end{equation}
The sectional curvature\textbf{ }$\mathcal{K}$ in Eq. (\ref{geo-deviation})
assumes a constant value throughout the manifold. In particular, when
$\mathcal{K}<0$, unstable solutions of Eq. (\ref{geo-deviation}) become%
\begin{equation}
J^{\mu}\left(  \xi\right)  =\frac{\omega_{0}^{\mu}}{\sqrt{-\mathcal{K}}}%
\sinh\left(  \sqrt{-\mathcal{K}}\xi\right)  \text{,}%
\end{equation}
with initial conditions $J^{\mu}\left(  0\right)  =0$ and $\frac{dJ^{\mu
}\left(  0\right)  }{d\xi}=\omega^{\mu}\left(  0\right)  =\omega_{0}^{\mu}%
\neq0$, respectively, for any $1\leq\mu\leq n$ with $n$ being the
dimensionality of the underlying manifold. For additional remarks concerning
the JLC equation, we suggest Refs. \cite{weinberg,carmo,ohanian}.

We point out that it would be intriguing to understand the behavior of the
Jacobi vector fields within the geometry of the Kaniadakis statistical
mechanics emerging from a one deformation parameter $\kappa$
\cite{kaniadakis02}. We leave this fascinating line of study to future
scientific inquiry. For a schematic description of the behavior of the IGE and
the Jacobi field for two-dimensional surfaces with distinct (Gaussian)
curvatures, we refer to Table I and Fig.\textbf{ }$1$.

In the next section, making use of the complexity quantifiers introduced in
Eqs. (\ref{IGE}), (\ref{Ricci}), and (\ref{Jintensity}), we present numerous
illustrative examples within the IGAC framework.

\begin{table}[t]
\centering
\begin{tabular}
[c]{|c|c|c|c|}\hline
Surface & Curvature & Jacobi Field Behavior & IGE Behavior\\\hline
sphere & positive & oscillatory & sublogarithmic\\
cylinder & zero & linear & logarithmic\\
hyperboloid & negative & exponential & linear\\\hline
\end{tabular}
\caption{Schematic description of the behavior of the IGE and the Jacobi field
for different types of two-dimensional surfaces characterized by distinct
constant values of their Gaussian curvature. For such surfaces, the sectional
and the scalar curvatures coincide, while the Gaussian curvature is simply
one-half of the scalar curvature. In particular, positive curvature causes
geodesics to converge while negative curvature causes geodesics to spread out.
More specifically, in flat, positively, and negatively curved manifolds, the
geodesic deviation equation yields deviations of nearby geodesics that exhibit
linear, oscillatory, and exponential behaviors, respectively. Moreover, the
volumes of the manifolds regions explored during the entropic motion tend to
increase while transitioning from positively to negatively curved manifolds.
Correspondingly, the IGE exhibits its maximum growth (that is, linear growth)
in the presence of exponential instability on negatively curved manifolds.}%
\end{table}

\begin{figure}[t]
\centering
\includegraphics[width=1\textwidth] {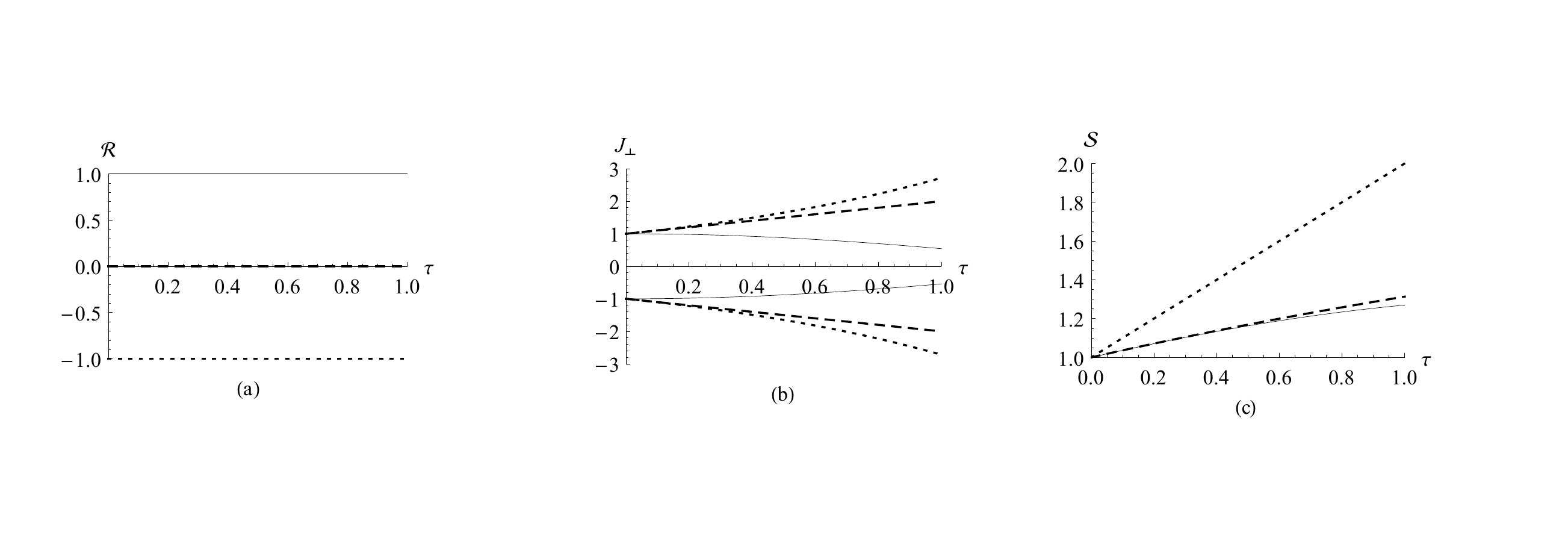}\caption{Graphical depictions of the
links among curvature, Jacobi fields, and IGE. In (a), we depict the constant
scalar curvature of a positively curved manifold (solid line), a flat manifold
(dashed line), and a negatively curved manifold (dotted line). In (b), we
illustrate the behavior of the normal components of the Jacobi fields
quantifying how nearby geodesics are changing in the normal direction (that
is, the direction that is orthogonal to the unit tangent vector of the
geodesic) as we move along the geodesics. In the positive, flat, and negative
curvature cases, we observe oscillatory behavior (solid line), linear behavior
(dashed line), and exponential behavior (dotted line), respectively. Finally,
in (c) we plot the temporal behavior of the IGE in the positive
(sublogarithmic behavior, solid line), flat (logarithmic behavior, dashed
line), and negative (linear behavior, dotted line) curvature cases.}%
\end{figure}

\section{Applications}

In this section, with the help of the three complexity quantifiers introduced
above, we report the results of several applications of the IGAC in which the
complexity of geodesic trajectories on statistical manifolds are quantified.
We present these illustrative examples in a chronological order, from the
first one to the last one. For brevity, we omit technical details and confine
the presentation to our own information geometric approach to complexity.
Early notions and applications of the IGAC originally appeared in Refs.
\cite{cafaroaip06, cafaroaip07, cafaroaip08}. For a recent review of the IGAC
framework, we refer to Refs. \cite{ali18,ali17,alips12, cafarobrescia13} and
Ref. \cite{cafaroamc10}, respectively.

\subsection{Uncorrelated Gaussian statistical models}

In \cite{cafaropd07, cafaroijtp08}, the IGAC framework was employed to study
the information geometric features of a system of arbitrary nature,
characterized by $l$ degrees of freedom. Each of these degrees of freedom is
described by two relevant pieces of information, namely its mean and variance.
The infinitesimal line element for this model is given by \cite{cafaroijtp08},%
\begin{equation}
ds^{2}\overset{\text{def}}{=}\sum_{k=1}^{l}\frac{1}{\sigma_{k}^{2}}d\mu
_{k}^{2}+\frac{2}{\sigma_{k}^{2}}d\sigma_{k}^{2}\text{,}%
\end{equation}
with $\mu_{k}$ and $\sigma_{k}$ denoting the expectation value and the square
root of the variance of the microvariable $x_{k}$, respectively. It was found
that the family of statistical models associated to such a system is Gaussian
in form. Specifically, it was determined that\textbf{ }this set of Gaussian
distributions yields a non-maximally symmetric $2l$-dimensional statistical
manifold $\mathcal{M}_{s}$ whose scalar curvature $\mathcal{R}_{\mathcal{M}%
_{s}}$ assumes a constant negative value that is\textbf{ }proportional to the
number of degrees of freedom of the system,%
\begin{equation}
\mathcal{R}_{\mathcal{M}_{s}}=-l\text{.} \label{curvature18}%
\end{equation}
It was determined that the system explores volume elements on $\mathcal{M}%
_{s}$ at an exponential rate. In particular, the IGE $\mathcal{S}%
_{\mathcal{M}_{s}}$\ was found to increase in a linear fashion in the
asymptotic temporal limit (more precisely, in asymptotic limit of the
statistical affine parameter $\tau$) and is proportional to the number of
degrees of freedom $l$,%
\begin{equation}
\mathcal{S}_{\mathcal{M}_{s}}\left(  \tau\right)  \overset{\tau\rightarrow
\infty}{\sim}l\lambda\tau\text{.} \label{entropy18}%
\end{equation}
The quantity $\lambda$ in Eq. (\ref{entropy18}) denotes the maximal positive
Lyapunov exponent that specifies the statistical model. Geodesic trajectories
on $\mathcal{M}_{s}$ were found to be hyperbolic curves. Finally, it was
determined\textbf{ }that in the asymptotic limit, the Jacobi vector field
intensity $J_{\mathcal{M}_{s}}$ is exponentially divergent and is proportional
to the number of degrees of freedom $l$,%
\begin{equation}
J_{\mathcal{M}_{s}}\left(  \tau\right)  \overset{\tau\rightarrow\infty}{\sim
}l\exp\left(  \lambda\tau\right)  \text{.} \label{jacobi18}%
\end{equation}
Given that the exponential divergence of the Jacobi vector field intensity
$J_{\mathcal{M}_{s}}$ is an established classical feature of chaos, based on
the results displayed in Eqs. (\ref{curvature18}), (\ref{entropy18}) and
(\ref{jacobi18}), the authors suggest that $\mathcal{R}_{\mathcal{M}_{s}}$,
$\mathcal{S}_{\mathcal{M}_{s}}$ and $J_{\mathcal{M}_{s}}$ each behave as
legitimate measures of chaoticity, with each indicator being proportional to
the number of Gaussian-distributed microstates of the system.\textbf{
}Although this result was verified in the context of this special scenario,
the proportionality among $\mathcal{R}_{\mathcal{M}_{s}}$, $\mathcal{S}%
_{\mathcal{M}_{s}}$ and $J_{\mathcal{M}_{s}}$ constitutes the first known
example appearing in the literature of a possible connection among information
geometric indicators of chaoticity obtained from probabilistic modeling of
dynamical systems.\textbf{ }In \ this first example, we have compared all
three measures\textbf{ }$\mathcal{R}_{\mathcal{M}_{s}}$\textbf{, }%
$\mathcal{S}_{\mathcal{M}_{s}}$\textbf{ }and\textbf{ }$J_{\mathcal{M}_{s}}%
$\textbf{. }Although we have not performed such a comparative analysis in all
applications, we shall attempt to mention curvature and/or Jacobi vector field
intensity behaviors whenever possible. Our emphasis here is especially on our
entropic measure of complexity. For more details on the other types of
complexity indicators, we refer to our original works cited in this manuscript.

\subsection{Correlated Gaussian statistical models}

In \cite{aliphysica10}, the IGAC framework was used to analyze the information
constrained dynamics of a system comprised of two correlated,
Gaussian-distributed microscopic degrees of freedom each having the same
variance. The infinitesimal line element for this model is given by
\cite{aliphysica10},%
\begin{equation}
ds^{2}\overset{\text{def}}{=}\frac{1}{\sigma^{2}}\left[  \frac{1}{1-r^{2}}%
d\mu_{x}^{2}+\frac{1}{1-r^{2}}d\mu_{y}^{2}-\frac{2r}{1-r^{2}}d\mu_{x}d\mu
_{y}+4d\sigma^{2}\right]  \text{,} \label{linecor}%
\end{equation}
with $\mu_{x}$ and $\mu_{y}$ denoting the expectation values of the
microvariables $x$ and $y$. The quantity $\sigma^{2}$, instead, is the
variance while $r$ is the usual correlation coefficient between $x$ and $y$.
The scalar curvature\textbf{ }$\mathcal{R}_{\mathcal{M}_{s}}$\textbf{ }of the
manifold with line element in Eq.\textbf{ }(\ref{linecor}) is\textbf{
}$\mathcal{R}_{\mathcal{M}_{s}}=-3/2$\textbf{. }The inclusion of microscopic
correlations give rise to asymptotic compression of the statistical
macrostates explored by the system at a faster rate than that observed in the
absence of microscopic correlations. Specifically, it was determined that in
the asymptotic limit%
\begin{equation}
\left[  \exp(\mathcal{S}_{\mathcal{M}_{s}}\left(  \tau\right)  )\right]
_{\text{correlated}}\overset{\tau\rightarrow\infty}{\sim}\mathcal{F}\left(
r\right)  \cdot\left[  \exp(\mathcal{S}_{\mathcal{M}_{s}}\left(  \tau\right)
)\right]  _{\text{uncorrelated}}\text{,} \label{fuckyou}%
\end{equation}
where the function $\mathcal{F}\left(  r\right)  $ in Eq. (\ref{fuckyou})
with\textbf{ }$0\leq$ $\mathcal{F}\left(  r\right)  \leq1$ is defined as
\cite{aliphysica10},%
\begin{equation}
\mathcal{F}\left(  r\right)  \overset{\text{def}}{=}\frac{1}{2^{\frac{5}{2}}%
}\left[  \sqrt{\frac{4\left(  4-r^{2}\right)  }{\left(  2-2r^{2}\right)  ^{2}%
}}\left(  \frac{2+r}{4\left(  1-r^{2}\right)  }\right)  ^{-\frac{3}{2}%
}\right]  \text{.}%
\end{equation}
The function $\mathcal{F}\left(  r\right)  $ is a monotone decreasing
compression envelope $\forall r\in\left(  0,1\right)  $. This result provides
an explicit link between correlations at the \emph{microscopic} level and
complexity at the \emph{macroscopic} level. It also furnishes a transparent
and concise description of the functional change of the macroscopic complexity
of the underlying statistical manifold caused by the occurrence of microscopic correlations.

\subsection{Inverted harmonic oscillators}

Generally speaking, the fundamental issues addressed by the General Theory of
Relativity are twofold: firstly, one wishes to understand how the geometry of
space-time evolves in response to the presence of mass-energy distributions;
secondly, one seeks to investigate how configurations of mass-energy move in
dynamical space-time\ geometry. By contrast, within the IGAC framework, one is
concerned only with the manner in which systems move within a given
statistical geometry, while the evolution of the statistical manifold itself
is neglected. The recognition that there exist two separate and distinct
characteristics to consider regarding the interplay between mass-energy and
space-time geometry served as a catalyst in the development of the IGAC
framework, ultimately leading to a rather interesting finding. The first
result obtained in this novel research direction was proposed by Caticha and
Cafaro in \cite{catichaaip07}. In that article, the possibility of utilizing
well established principles of inference to obtain Newtonian dynamics from
relevant prior information encoded in a suitable statistical manifold was
investigated. The primary working assumption in that derivation was the
assumed existence of an irreducible uncertainty in the location of particles.
This uncertainty requires the state of a particle to be described by a
probability distribution. The resulting configuration space is therefore a
statistical manifold whose Riemannian geometry is specified by the Fisher-Rao
information metric. The expected trajectory is a consequence of the MrE
method, with the latter being regarded as a principle of inference. An
unexpected consequence of this approach is that no additional physical
postulates such as an equation of motion, principle of least action, nor the
concept of momentum, mass, phase space or external time are required. Newton's
mechanics involving any number of self-interacting particles as well as
particles interacting with external fields is entirely recovered by the
resulting entropic dynamics. Indeed, a powerful result of this approach is the
fact that interactions among particles as well as particle masses are all
justified in terms of the underlying statistical manifold.

Our next example will be of a more applied nature. In \cite{zurek94,zurek95},
Zurek and Paz explored the effects of decoherence in quantum chaos by
analyzing a single unstable harmonic oscillator with frequency $\Omega$ and
potential $V\left(  x\right)  $,%
\begin{equation}
V\left(  x\right)  \overset{\text{def}}{=}-\frac{\Omega^{2}x^{2}}{2}\text{,}
\label{vimo}%
\end{equation}
coupled to an external environment. They determined that in the reversible
classical limit, the von Neumann entropy of such a system increases linearly
at a rate determined by the Lyapunov exponent $\Omega$ according to,
\begin{equation}
\mathcal{S}_{\text{quantum}}^{\text{(chaotic)}}\left(  \tau\right)
\overset{\tau\rightarrow\infty}{\sim}\Omega\tau\text{.} \label{hud1}%
\end{equation}
Building upon the results obtained in \cite{catichaaip07}, an information
geometric analogue of the Zurek-Paz quantum chaos criterion in the classical
reversible limit was proposed in \cite{cafaroejtp08,cafarocsf09}. In these
works, the IGAC\ framework was employed to study a set of\ $l$,
three-dimensional, anisotropic, uncoupled, inverted harmonic oscillators (IHO)
with an Ohmic distributed frequency spectrum. In this example, the
infinitesimal line element is given by%
\begin{equation}
ds^{2}\overset{\text{def}}{=}\left[  1-\Phi\left(  \theta\right)  \right]
\delta_{\mu\nu}\left(  \theta\right)  d\theta^{\mu}d\theta^{\nu}\text{,}%
\end{equation}
where $\Phi\left(  \theta\right)  $ is defined as,%
\begin{equation}
\Phi\left(  \theta\right)  =\sum_{k=1}^{l}u_{k}\left(  \theta\right)  \text{,}%
\end{equation}
with $u_{k}\left(  \theta\right)  \overset{\text{def}}{=}-(1/2)\omega_{k}%
^{2}\theta_{k}^{2}$ and $\omega_{k}$ being the frequency of the $k$-th
inverted harmonic oscillator\textbf{.} Neglecting mathematical details, it was
demonstrated in Refs. \cite{cafaroejtp08,cafarocsf09} that the asymptotic
temporal behavior of the IGE for such a system becomes,%
\begin{equation}
\mathcal{S}_{\mathcal{M}_{\text{IHO}}^{\text{(}l\text{)}}}\left(  \tau\text{;
}\omega_{1}\text{,\ldots, }\omega_{l}\right)  \overset{\tau\rightarrow\infty
}{\sim}\Omega\tau\text{,} \label{Fin}%
\end{equation}
where,
\begin{equation}
\Omega\overset{\text{def}}{=}\overset{l}{\underset{i=1}{\sum}}\omega
_{i}\text{,}%
\end{equation}
and $\omega_{i}$ with $1\leq i\leq l$ is the frequency of the $i^{\text{th}}$
IHO. Equation (\ref{Fin}) indicates an asymptotic, linear IGE growth for the
set of IHOs and can be regarded as an extension of the result of Zurek and Paz
appearing in Refs. \cite{zurek94,zurek95} to an ensemble of anisotropic,
uncoupled, inverted harmonic oscillators in the context of the IGAC. We remark
that Eq. (\ref{Fin}) was proposed as the classical IG analogue of Eq.
(\ref{hud1}) in Refs. \cite{cafaroejtp08,cafarocsf09}.

\subsection{Quantum spin chains}

In \cite{cafaromplb08,cafarophysica08}, the IGAC\ was used to study the ED on
statistical manifolds whose elements are classical probability distribution
functions routinely employed in the study of regular and chaotic quantum
energy level statistics. Specifically, an IG description of the chaotic
(integrable) energy level statistics of a quantum antiferromagnetic Ising spin
chain immersed in a tilted (transverse) external magnetic field was presented.
The IGAC of a Poisson distribution coupled to an Exponential bath (that
specifies a spin chain in a \textit{transverse} magnetic field and corresponds
to the integrable case) along with that of a Wigner-Dyson distribution coupled
to a Gaussian bath (that specifies a spin chain in a \textit{tilted} magnetic
field and corresponds to the chaotic case) were investigated. The line
elements in the integrable and chaotic cases are given by\textbf{,}%
\begin{equation}
ds_{\mathrm{integrable}}^{2}\overset{\text{def}}{=}ds_{\mathrm{Poisson}}%
^{2}+ds_{\text{\textrm{Exponential}}}^{2}=\frac{1}{\mu_{A}^{2}}d\mu_{A}%
^{2}+\frac{1}{\mu_{B}^{2}}d\mu_{B}^{2}\text{,} \label{qsc1}%
\end{equation}
and,%
\begin{equation}
ds_{\mathrm{chaotic}}^{2}\overset{\text{def}}{=}ds_{\mathrm{Wigner-Dyson}}%
^{2}+ds_{\text{\textrm{Gaussian}}}^{2}=\frac{4}{\mu_{A}^{\prime2}}d\mu
_{A}^{\prime2}+\frac{1}{\sigma_{B}^{\prime2}}d\mu_{B}^{\prime2}+\frac
{2}{\sigma_{B}^{\prime2}}d\sigma_{B}^{\prime2}\text{,} \label{qsc2}%
\end{equation}
respectively. In Eq. (\ref{qsc1}), $\mu_{A}$ and $\mu_{B}$ are the average
spacing of the energy levels and the average intensity of the magnetic field,
respectively. A similar notation is employed for the second scenario described
in Eq. (\ref{qsc2}) where, clearly, $\sigma_{B}^{\prime2}$ denotes the
variance of the intensity of the magnetic field.\textbf{ }Remarkably, it was
determined that in the former case, the IGE shows asymptotic logarithmic
growth,%
\begin{equation}
\mathcal{S}_{\mathcal{M}_{s}}^{\text{(integrable)}}\left(  \tau\right)
\overset{\tau\rightarrow\infty}{\sim}c\log\left(  \tau\right)  +\tilde
{c}\text{,} \label{sintegrable}%
\end{equation}
whereas in the latter case, the IGE shows asymptotic linear growth,%
\begin{equation}
\mathcal{S}_{\mathcal{M}_{s}}^{\text{(chaotic)}}\left(  \tau\right)
\overset{\tau\rightarrow\infty}{\sim}\mathcal{K}\tau\text{.} \label{schaos}%
\end{equation}
We emphasize that the quantities $c$ and $\tilde{c}$ in Eq. (\ref{sintegrable}%
) are integration constants that depend upon the dimensionality of the
statistical manifold and the boundary constraint conditions on the statistical
variables, respectively. The quantity $\mathcal{K}$ appearing in Eq.
(\ref{schaos}) denotes a model parameter describing the asymptotic temporal
rate of change of the IGE. The findings described above suggest that the IGAC
framework may prove useful in the analysis of applications involving quantum
energy level statistics. It is worth noting that in such cases, the
IGE\ effectively serves the role of the standard entanglement entropy used in
quantum information science \cite{prosenpre07,prosenpra07}.

\subsection{Statistical embedding and complexity reduction}

Expanding upon the analysis presented in Ref. \cite{cafarops10}, Cafaro and
Mancini utilized the IGAC framework in Ref. \cite{cafaropd11} to study the
$2l$-dimensional Gaussian statistical model $\mathcal{M}_{s}$ induced by an
appropriate embedding within a larger $4l$-dimensional Gaussian manifold. The
geometry of the $4l$-dimensional Gaussian manifold is defined by a Fisher-Rao
information metric $g_{\mu\nu}$ with non-vanishing off-diagonal elements. It
should be noted that these non-vanishing off-diagonal terms arise due to the
occurrence of macroscopic correlation coefficients $\rho_{k}$ with $1\leq
k\leq l$ that specify the embedding constraints among the statistical
variables on the larger manifold. The infinitesimal line element is given by
\cite{cafaropd11},%
\begin{equation}
ds^{2}\overset{\text{def}}{=}\sum_{k=1}^{l}\frac{1}{\sigma_{2k-1}^{2}}\left[
d\mu_{2k-1}^{2}+2\rho_{2k-1}d\mu_{2k-1}d\sigma_{2k-1}+2d\sigma_{2k-1}%
^{2}\right]  \text{,}%
\end{equation}
with $\rho_{2k-1}$ defined as,%
\begin{equation}
\rho_{2k-1}\overset{\text{def}}{=}\frac{\frac{\partial\mu_{2k}}{\partial
\mu_{2k-1}}\frac{\partial\mu_{2k}}{\partial\sigma_{2k-1}}}{\left[  1+\left(
\frac{\partial\mu_{2k}}{\partial\mu_{2k-1}}\right)  ^{2}\right]  ^{1/2}\left[
2+\frac{1}{2}\left(  \frac{\partial\mu_{2k}}{\partial\sigma_{2k-1}}\right)
^{2}\right]  ^{1/2}}\text{,}%
\end{equation}
where $\sigma_{2k}=\sigma_{2k-1}$ and $\mu_{2k}=\mu_{2k}\left(  \mu
_{2k-1}\text{, }\sigma_{2k-1}\right)  $ for any $1\leq k\leq l$. Two
significant results were obtained. First, a power law decay of the IGE at a
rate determined by the correlation coefficients $\rho_{k}$ was observed%
\begin{equation}
\mathcal{S}_{\mathcal{M}_{s}}\left(  \tau;l,\lambda_{k},\rho_{k}\right)
\overset{\tau\rightarrow\infty}{\sim}\log\left[  \Lambda\left(  \rho
_{k}\right)  +\frac{\tilde{\Lambda}\left(  \rho_{k},\lambda_{k}\right)  }%
{\tau}\right]  ^{l}\text{,} \label{peppino}%
\end{equation}
with $\rho_{k}=\rho_{s}$ $\forall k$ and $s=1,\ldots,l$, where
\begin{equation}
\Lambda\left(  \rho_{k}\right)  \overset{\text{def}}{=}\frac{2\rho_{k}%
\sqrt{2-\rho_{k}^{2}}}{1+\sqrt{\Delta\left(  \rho_{k}\right)  }}\text{,
}\tilde{\Lambda}\left(  \rho_{k},\lambda_{k}\right)  \overset{\text{def}}%
{=}\frac{\sqrt{\Delta\left(  \rho_{k}\right)  \left(  2-\rho_{k}^{2}\right)
}\log\left[  \Sigma\left(  \rho_{k},\lambda_{k},\alpha_{\pm}\right)  \right]
}{\rho_{k}\lambda_{k}}\text{, and }\alpha_{\pm}\left(  \rho_{k}\right)
\overset{\text{def}}{=}\frac{1}{2}\left(  3\pm\sqrt{\Delta\left(  \rho
_{k}\right)  }\right)  \text{.}%
\end{equation}
The quantity $\Sigma\left(  \rho_{k},\lambda_{k},\alpha_{\pm}\right)  $ is a
strictly positive function of its arguments for $0\leq\rho_{k}<1$ and
is\textbf{ }given by \cite{cafaropd11},%
\begin{equation}
\Sigma\left(  \rho_{k},\lambda_{k},\alpha_{\pm}\right)  \overset{\text{def}%
}{=}-\frac{\Xi_{k}}{4\lambda_{k}}\frac{1+\sqrt{\Delta\left(  \rho_{k}\right)
}}{1-\sqrt{\Delta\left(  \rho_{k}\right)  }}\sqrt{\frac{2\alpha_{-}\left(
\rho_{k}\right)  }{\alpha_{+}\left(  \rho_{k}\right)  }}\text{,}%
\end{equation}
where $\Xi_{k}$ and $\lambda_{k}$ are real, positive constants of integration,
and%
\begin{equation}
\Delta\left(  \rho_{k}\right)  \overset{\text{def}}{=}1+4\rho_{k}^{2}\text{.}%
\end{equation}
Equation (\ref{peppino}) represents the first main finding reported in Ref.
\cite{cafaropd11} and can be interpreted as a quantitative indication that the
IGC of a system decreases in response to the emergence of correlational
structures. Second, it was demonstrated that the presence of embedding
constraints among the Gaussian macrovariables of the larger $4l$-dimensional
manifold results in an attenuation of the asymptotic exponential divergence of
the Jacobi field intensity on the embedded $2l$-dimensional manifold.
Neglecting mathematical details, it was determined in Ref. \cite{cafaropd11}
that in the asymptotic limit $\tau\gg1$,%
\begin{equation}
0\leq\frac{J_{\mathcal{M}_{s}}^{2l\text{-embedded}}\left(  \tau\right)
}{J_{\mathcal{M}_{s}}^{4l\text{-larger}}\left(  \tau\right)  }<1\text{.}
\label{peppe2}%
\end{equation}
Equation (\ref{peppe2}) constitute the second main finding reported in Ref.
\cite{cafaropd11}. The observed attenuation of the asymptotic exponential
divergence of the Jacobi vector field associated with the larger
$4l$-manifold, suggests that the occurrence of such embedding constraint
relations results in an asymptotic compression of the macrostates explored on
the statistical manifold $\mathcal{M}_{s}$.\textbf{ }These two findings serve
to advance, in a\ non-trivial manner, the goal of developing a description of
complexity of either macroscopically or microscopically correlated,
multi-dimensional Gaussian statistical models relevant in the modelling of
complex systems.

\subsection{Entanglement induced via scattering}

Guided by the original study appearing in \cite{kim11}, the IGAC framework was
employed to furnish an IG viewpoint on the phenomena of quantum entanglement
emerging via\ $s$-wave scattering between interacting Gaussian wave packets in
Refs. \cite{kimpla11,kim12}. Within the IGAC framework, the pre and post
quantum scattering scenarios associated with elastic, head-on collision are
hypothesized to be macroscopic manifestations arising from underlying
microscopic statistical structures. By exploiting this working hypothesis, the
pre and post quantum scattering scenarios were modeled by uncorrelated and
correlated Gaussian statistical models, respectively. Using the standard
notation used so far in this article, the infinitesimal line elements in the
absence and presence of correlations are given by%
\begin{equation}
ds_{\mathrm{no}\text{-}\mathrm{correlations}}^{2}=\frac{1}{\sigma^{2}}\left[
d\mu_{x}^{2}+d\mu_{y}^{2}+4d\sigma^{2}\right]  \text{,} \label{lineaa}%
\end{equation}
\textbf{and,}%
\begin{equation}
ds_{\mathrm{correlations}}^{2}\overset{\text{def}}{=}\frac{1}{\sigma^{2}%
}\left[  \frac{1}{1-r^{2}}d\mu_{x}^{2}+\frac{1}{1-r^{2}}d\mu_{y}^{2}-\frac
{2r}{1-r^{2}}d\mu_{x}d\mu_{y}+4d\sigma^{2}\right]  \text{,} \label{lineaaa}%
\end{equation}
respectively. The scalar curvature\textbf{ }$\mathcal{R}_{\mathcal{M}_{s}}%
$\textbf{ }of the manifolds with line elements in Eqs. (\ref{lineaa}) and
(\ref{lineaaa}) is\textbf{ }$\mathcal{R}_{\mathcal{M}_{s}}=-3/2$\textbf{.
}Using such a hybrid modeling approach enabled the authors to express the
entanglement strength in terms of the scattering potential and incident
particle energy. Moreover, the manner in which the entanglement duration is
related\ to the scattering potential and incident particle energy was
furnished with a possible explanation. Finally, the link between complexity of
informational geodesic paths and entanglement was discussed. In particular, it
was demonstrated that in the asymptotic limit,%
\begin{equation}
\left[  \exp(\mathcal{S}_{\mathcal{M}_{s}}\left(  \tau\right)  )\right]
_{\text{correlated}}\overset{\tau\rightarrow\infty}{\sim}\mathcal{F}\left(
r\right)  \cdot\left[  \exp(\mathcal{S}_{\mathcal{M}_{s}}\left(  \tau\right)
)\right]  _{\text{uncorrelated}}\text{,} \label{family}%
\end{equation}
where the function $\mathcal{F}\left(  r\right)  $ in Eq. (\ref{family}) with
$0\leq\mathcal{F}\left(  r\right)  \leq1$ is defined as,%
\begin{equation}
\mathcal{F}\left(  r\right)  \overset{\text{def}}{=}\sqrt{\frac{1-r}{1+r}%
}\text{.}%
\end{equation}
The function $\mathcal{F}\left(  r\right)  $ is a monotone decreasing
compression factor with $0<r<1$. The analysis proposed in Refs.
\cite{kimpla11,kim12} is a significant progress toward the understanding among
the concepts of entanglement and statistical micro-correlations, as well as
the impact of micro-correlations on the complexity of informational geodesic
paths. The finding appearing in Eq. (\ref{family}) suggest that the IGAC
construct may prove useful in developing a sound IG perspective of the
phenomenon of quantum entanglement.

\subsection{Softening of classical chaos by quantization}

Expanding upon the original analysis presented in Refs.
\cite{cafaroaip12,aliaip12,giffinaip13}, the IGAC framework was utilized to
investigate the entropic dynamics and information geometry of a
three-dimensional Gaussian statistical model as well as the two-dimensional
Gaussian statistical model derived from the former model by introducing the
following macroscopic information constraint,%
\begin{equation}
\sigma_{x}\sigma_{y}=\Sigma^{2}\text{,} \label{macro}%
\end{equation}
where $\Sigma^{2}\in%
\mathbb{R}
_{0}^{+}$. The quantities $x$ and $y$ label the microscopic degrees of freedom
of the system. The constraint equation (\ref{macro}) resembles the standard
minimum uncertainty relation encountered in quantum mechanics
\cite{cafaroosid12}. The infinitesimal line elements in the $3D$- and
$2D$-Gaussian statistical models are given by%
\begin{equation}
ds_{3D}^{2}\overset{\text{def}}{=}\frac{1}{\sigma_{x}^{2}}d\mu_{x}^{2}%
+\frac{2}{\sigma_{x}^{2}}d\sigma_{x}^{2}+\frac{2}{\sigma_{y}^{2}}d\sigma
_{y}^{2}\text{,} \label{3DD}%
\end{equation}
and,%
\begin{equation}
ds_{2D}^{2}\overset{\text{def}}{=}\frac{1}{\sigma^{2}}d\mu_{x}^{2}+\frac
{4}{\sigma^{2}}d\sigma^{2}\text{,} \label{2DD}%
\end{equation}
respectively. Note that the expectation value $\mu_{y}$ of the microvariable
$y$ is set equal to zero in Eq. (\ref{3DD}), while $\sigma_{x}=\sigma$ with
$\sigma_{x}\sigma_{y}=\Sigma^{2}$ in Eq. (\ref{2DD}). Furthermore, the scalar
curvatures corresponding to the $3D$ and $2D$ cases are equal to\textbf{
}$\mathcal{R}_{3D}=-1$\textbf{ }and\textbf{ }$\mathcal{R}_{2D}=-1/2$\textbf{,
}respectively. It was determined that the complexity of the $2D$-Gaussian
statistical model specified by the IGE is relaxed when compared with the
complexity of the $3D$-Gaussian statistical model,%
\begin{equation}
\mathcal{S}_{\mathcal{M}_{s}}^{\text{(}2D\text{)}}\left(  \tau\right)
\overset{\tau\rightarrow\infty}{\sim}\left(  \frac{\lambda_{2D}}{\lambda_{3D}%
}\right)  \cdot\mathcal{S}_{\mathcal{M}_{s}}^{\text{(}3D\text{)}}\left(
\tau\right)  \text{,}%
\end{equation}
with $\lambda_{2D}$ and $\lambda_{3D}$ being both positive model parameters
(satisfying the condition $\lambda_{2D}\leq\lambda_{3D}$) that express the
asymptotic temporal rates of change of the IGE in the $2D$ and $3D$ cases,
respectively. Motivated by the connection between the macroscopic information
constraint (\ref{macro}) on the variances and the phase-space coarse-graining
due to the Heisenberg uncertainty relations, the authors suggest their work
may shed light on the phenomenon of classical chaos suppression arising from
the process of quantization when expressed in an IG setting. It is worth
noting that a similar analysis was implemented in Ref. \cite{giffinentropy13}
where the work in Ref. \cite{cafaroosid12} was generalized to a scenario where
- in conjunction with the macroscopic constraint in Eq. (\ref{macro}) - the
microscopic degrees of freedom $x$ and $y$ of the system are also correlated.

\subsection{Topologically distinct correlational structures}

In Ref. \cite{felice14}, \ the asymptotic behavior of the IGE associated with
either bivariate or trivariate Gaussian statistical models, with or without
micro-correlations, was analyzed by Felice and coworkers. For correlated
cases, several correlational configurations among the microscopic degrees of
freedom of the system were taken into consideration. It was found that the
complexity of macroscopic inferences is dependent on the quantity of
accessible microscopic information, as well as on how such microscopic
information are correlated. Specifically, in the mildly connected case defined
by a trivariate statistical model with two correlations among the three
degrees of freedom of the system, the infinitesimal line element is%
\begin{equation}
\left(  \left[  ds^{2}\right]  _{\text{trivariate}}^{\text{(mildly
connected)}}\right)  _{\text{correlated}}\overset{\text{def}}{=}\frac
{1}{\sigma^{2}}\frac{3-4r}{1-2r^{2}}d\mu^{2}+\frac{6}{\sigma^{2}}d\sigma
^{2}\text{.}\label{fu1}%
\end{equation}
Moreover, the infinitesimal line element in the uncorrelated trivariate case
is given by%
\begin{equation}
\left(  \left[  ds^{2}\right]  _{\text{trivariate}}\right)
_{\text{uncorrelated}}\overset{\text{def}}{=}\frac{3}{\sigma^{2}}d\mu
^{2}+\frac{6}{\sigma^{2}}d\sigma^{2}\text{.}\label{fu2}%
\end{equation}
In Eqs. (\ref{fu1}) and (\ref{fu2}), $\mu$, $\sigma$, and $r$ denote the
expectation value, the standard deviation, and the correlation coefficient,
respectively. It was determined that in the asymptotic limit,
\begin{equation}
\left(  \exp\left[  \mathcal{S}_{\text{trivariate}}^{\text{(mildly
connected)}}\left(  \tau\right)  \right]  \right)  _{\text{correlated}%
}\overset{\tau\rightarrow\infty}{\sim}\mathcal{\tilde{R}}_{\text{trivariate}%
}^{\text{(mildly connected)}}\left(  r\right)  \left(  \exp\left[
\mathcal{S}_{\text{trivariate}}^{\text{(mildly connected)}}\left(
\tau\right)  \right]  \right)  _{\text{uncorrelated}}\text{,}\label{cacchio2}%
\end{equation}
where%
\begin{equation}
\mathcal{\tilde{R}}_{\text{trivariate}}^{\text{(mildly connected)}}\left(
r\right)  \overset{\text{def}}{=}\sqrt{\frac{3\left(  1-2r^{2}\right)  }%
{3-4r}}\text{.}%
\end{equation}
In Eq. (\ref{cacchio2}), the quantity $r$ is the micro-correlation
coefficient. The function $\mathcal{\tilde{R}}_{\text{trivariate}%
}^{\text{(mildly connected)}}\left(  r\right)  $ shows non-monotone behavior
in the correlation parameter $r$ and assumes a value of zero at the extrema of
the permitted range $r\in\left(  -\sqrt{2}/2\text{, }\sqrt{2}/2\right)  $. By
contrast, for closed bivariate configurations where all microscopic variables
are correlated with each other, the complexity ratio between correlated and
uncorrelated cases presents monotone behavior in the correlation parameter
$r$. For example, in the fully connected bivariate Gaussian case with $\mu
_{x}=\mu_{y}=\mu$ and $\sigma_{x}=\sigma_{y}=\sigma$, the infinitesimal line
element is%
\begin{equation}
\left(  \left[  ds^{2}\right]  _{\text{bivariate}}^{\text{(fully connected)}%
}\right)  _{\text{correlated}}\overset{\text{def}}{=}\frac{2}{\sigma^{2}}%
\frac{1}{1+r}d\mu^{2}+\frac{4}{\sigma^{2}}d\sigma^{2}\text{.}%
\end{equation}
It was found that%
\begin{equation}
\left(  \exp\left[  \mathcal{S}_{\text{bivariate}}^{\text{(fully connected)}%
}\left(  \tau\right)  \right]  \right)  _{\text{correlated}}\overset
{\tau\rightarrow\infty}{\sim}\mathcal{\tilde{R}}_{\text{bivariate}%
}^{\text{(fully connected)}}\left(  r\right)  \left(  \exp\left[
\mathcal{S}_{\text{bivariate}}^{\text{(fully connected)}}\left(  \tau\right)
\right]  \right)  _{\text{uncorrelated}}\text{,}%
\end{equation}
where%
\begin{equation}
\mathcal{\tilde{R}}_{\text{bivariate}}^{\text{(fully connected)}}\left(
r\right)  \overset{\text{def}}{=}\sqrt{1+r}\text{.}%
\end{equation}
Finally, in the fully connected trivariate Gaussian case with trivariate
models having all microscopic variables correlated with each other, the
infinitesimal line element is%
\begin{equation}
\left(  \left[  ds^{2}\right]  _{\text{trivariate}}^{\text{(fully connected)}%
}\right)  _{\text{correlated}}\overset{\text{def}}{=}\frac{3}{\sigma^{2}}%
\frac{1}{1+2r}d\mu^{2}+\frac{6}{\sigma^{2}}d\sigma^{2}\text{.}%
\end{equation}
It was determined in this case that%
\begin{equation}
\left(  \exp\left[  \mathcal{S}_{\text{trivariate}}^{\text{(fully connected)}%
}\left(  \tau\right)  \right]  \right)  _{\text{correlated}}\overset
{\tau\rightarrow\infty}{\sim}\mathcal{\tilde{R}}_{\text{trivariate}%
}^{\text{(fully connected)}}\left(  r\right)  \left(  \exp\left[
\mathcal{S}_{\text{trivariate}}^{\text{(fully connected)}}\left(  \tau\right)
\right]  \right)  _{\text{uncorrelated}}\text{,}\label{cacchio}%
\end{equation}
where%
\begin{equation}
\mathcal{\tilde{R}}_{\text{trivariate}}^{\text{(fully connected)}}\left(
r\right)  \overset{\text{def}}{=}\sqrt{1+2r}\text{.}%
\end{equation}
These results imply that in the fully connected bivariate and trivariate
configurations, the ratios $\mathcal{\tilde{R}}_{\text{bivariate}%
}^{\text{(fully connected)}}\left(  r\right)  $ and $\mathcal{\tilde{R}%
}_{\text{trivariate}}^{\text{(fully connected)}}\left(  r\right)  $ both
present monotone behavior in $r$ over the open intervals $\left(  -1,1\right)
$ and $\left(  -1/2,1\right)  $, respectively. On the other hand, in the
mildly connected trivariate scenario appearing in Eq. (\ref{cacchio2}), an
extrema in the function $\mathcal{\tilde{R}}_{\text{trivariate}}%
^{\text{(mildly connected)}}\left(  r\right)  $ occurs at $r_{\text{peak}%
}=1/2$ $\geq0$. Such distinctly different behavior between mildly and fully
connected trivariate configurations can be attributed to the fact that when
making statistical inferences subject to the hypothesis of three positively
correlated Gaussian random variables, the system becomes frustrated because
the maximum entropy favorable state - characterized by minimum complexity - is
incompatible with the initial working hypothesis. Guided by these results, it
was suggested in Ref. \cite{felice14} that the impossibility of realizing the
maximally favorable state for specific correlational configurations among
microscopic degrees of freedom, viewed from an entropic inference perspective,
yields an information geometric analogue of the statistical physics
frustration effect that arise when loops are present \cite{sadoc06}.

\begin{table}[t]
\centering
\begin{tabular}
[c]{|c|c|c|}\hline
Math \& IGAC & Classical \& IGAC & Quantum \& IGAC\\\hline
Micro and macro correlations & Geometrization of Newtonian mechanics & Spin
chains and energy levels statistics\\
Statistical embeddings & Inverted harmonic oscillators & Scattering induced
entanglement\\
Topology and correlational structures & Macro effects from micro information &
Softening chaoticity by quantization\\\hline
\end{tabular}
\caption{Schematic description of existing mathematical, classical, and
quantum investigations within the IGAC.}%
\end{table}

\section{Final Remarks}

In this paper, we discussed the primary results obtained by the authors and
colleagues over an extended period of work on the IGAC framework. A summary of
the IGAC applications can be found in Table II. For ease of readability, we
have chosen to omit technicalities in our discussion. We are aware of several
unresolved issues within the IGAC framework, including a deep understanding of
the foundational aspects of the IGE measure of complexity. Further
developments of the framework are necessary, especially within a fully quantum
mechanical setting. For a more detailed list on limitations and future
directions of the IGAC approach, we refer the interested reader to Ref.
\cite{ali17}. In particular, we mentioned there that one of our main
objectives in the near future is to extend our comprehension of the
relationship between the IGE\ and the Kolmogorov-Sinai dynamical entropy
\cite{greven}, the coarse-grained Boltzmann entropy \cite{greven} and the von
Neumann entropy \cite{peres}, depending upon the peculiarity of the system
being investigated. Despite its limitations, we are pleased\ that our
theoretical modeling approach is steadily gaining interest in the community of
researchers. Indeed, there appears to be an increasing number of scientists
who either actively use, or who's work is linked to the theoretical framework
described in the present brief feature review article \cite{peng, peng2, r1,
r2, r3, FMP, r5, r6, r7, r8, r9, r10, r11, r12, r13, r14, r15, r16, r17,r19,
r18,r18a,steven,gomez20,summers20,desh21}.

\bigskip

\begin{acknowledgments}
C. C. acknowledges the hospitality of the \emph{United States Air Force
Research Laboratory} in Rome where part of his initial contribution to this
work was completed.
\end{acknowledgments}


\begin{thebibliography}{99}                                                                                               %


\bibitem {caticha12}A. Caticha, \emph{Entropic Inference and the Foundations
of Physics}; USP Press: S\~{a}o Paulo, Brazil, 2012; Available online: http://www.albany.edu/physics/ACaticha-EIFP-book.pdf.

\bibitem {amari}S. Amari and \ H. Nagaoka, \emph{Methods of Information
Geometry}, Oxford University Press (2000).

\bibitem {cafarophd}C. Cafaro, \emph{The Information Geometry of Chaos}, Ph.D.
thesis, State University of New York at Albany (2008).

\bibitem {ali18}S. A. Ali, C. Cafaro, S. Gassner, and A. Giffin, \emph{An
information geometric perspective on the complexity of macroscopic predictions
arising from incomplete information}, Adv. Math. Phys., Volume 2018, Article
ID 2048521 (2018).

\bibitem {felice18}D. Felice, C. Cafaro, and S. Mancini, \emph{Information
geometric methods for complexity}, Chaos \textbf{28}, 032101 (2018).

\bibitem {catichaED}A. Caticha, \emph{Entropic dynamics}, AIP Conf. Proc.
\textbf{617}, 302 (2002).

\bibitem {cafaropre16}C. Cafaro and S. A. Ali, \emph{Maximum caliber inference
and the stochastic Ising model}\textbf{, }Phys. Rev. \textbf{E94} 052145 (2016).

\bibitem {ali17}S. A.\ Ali and C. Cafaro, \emph{Theoretical investigations of
an information geometric approach to complexity}, Rev. Math. Phys.
\textbf{29}, 1730002 (2017).

\bibitem {cafaro07A}C. Cafaro and S. A. Ali, \emph{The spacetime algebra
approach to massive classical electrodynamics with magnetic monopoles},
Advances in Applied Clifford Algebras \textbf{17}, 23 (2007); C. Cafaro,
\emph{Finite-range electromagnetic interaction and magnetic charges: spacetime
algebra or algebra of physical space?}, Advances in Applied Clifford Algebras
\textbf{17}, 617 (2007).

\bibitem {cafaroamc10}C. Cafaro, A. Giffin, S. A. Ali, and D.-H. Kim,
\emph{Reexamination of an information geometric construction of entropic
indicators of complexity}, Appl. Math. Comput. \textbf{217}, 2944
(2010).\emph{\ }

\bibitem {kittel}C. Kittel, \emph{Elementary Statistical Physics}, John Wiley
\& Sons, Inc. (1958).

\bibitem {ito20}S. Ito, M. Oizumi, and S. Amari, \emph{Unified framework for
the entropy production and the stochastic interaction based on information
geometry}, Phys. Rev. Research \textbf{2}, 033048 (2020).

\bibitem {kaniadakis02}G. Kaniadakis, \emph{Statistical mechanics in the
context of special relativity}, Phys. Rev. \textbf{E66}, 056125 (2002).

\bibitem {fisher}R.A. Fisher, \emph{Theory of statistical estimation}, Proc.
Cambridge Philos. Soc. \textbf{122}, 700 (1925).

\bibitem {rao}C.R. Rao, \emph{Information and accuracy attainable in the
estimation of statistical parameters}, Bull. Calcutta Math. Soc. \textbf{37},
81 (1945).

\bibitem {cencov}N. N. Cencov, \emph{Statistical decision rules and optimal
inference}, Transl. Math. Monographs, vol. \textbf{53}, Amer. Math. Soc.,
Providence-RI (1981).

\bibitem {campbell}L. L. Campbell, \emph{An extended Cencov characterization
of the information metric}, Proc. Am. Math. Soc. \textbf{98}, 135 (1986).

\bibitem {weinberg}S. Weinberg, \emph{Gravitation and Cosmology}, John Wiley
\& Sons, Inc. (1972).

\bibitem {lee}J. M. Lee, \emph{Riemannian Manifolds: An Introduction to
Curvature}, Springer (1997).

\bibitem {cafaropd07}C. Cafaro and S. A. Ali, \emph{Jacobi fields on
statistical manifolds of negative curvature}, Physica \textbf{D70}, 234 (2007).

\bibitem {ohanian}H. C. Ohanian and R. Ruffini, \emph{Gravitation and
Spacetime}, W. W. Norton \& Company (1994).

\bibitem {carmo}M. P. do Carmo, \emph{Riemannian Geometry}, Birkhauser (1992).

\bibitem {cafaroaip06}C. Cafaro, S. A. Ali, and A. Giffin, \emph{An
application of reversible entropic dynamics on curved statistical manifolds},
AIP Conf. Proc. \textbf{872}, 243 (2006).

\bibitem {cafaroaip07}C. Cafaro, \emph{Information geometry and chaos on
negatively curved statistical manifolds}, AIP Conf. Proc. \textbf{954}, 175 (2007).

\bibitem {cafaroaip08}C. Cafaro, \emph{Recent theoretical progress on an
information geometrodynamical approach to chaos}, AIP Conf. Proc.
\textbf{1073}, 16 ( 2008).

\bibitem {alips12}S. A. Ali, C. Cafaro, A. Giffin, and D.-H. Kim,
\emph{Complexity characterization in a probabilistic approach to dynamical
systems through information geometry and inductive inference}, Physica Scripta
\textbf{85}, 025009 (2012).

\bibitem {cafarobrescia13}C. Cafaro,\emph{ Information geometric complexity of
entropic motion on curved statistical manifolds}. In Proceedings of the 12th
Joint European Thermodynamics Conference, Brescia, Italy, 1--5 July 2013;
Cartolibreria Snoopy: Brescia, Italy, 2013; Pilotelli, M., Beretta, G.P.,
Eds.; pp. 110--118.

\bibitem {cafaroijtp08}C. Cafaro, \emph{Information-geometric indicators of
chaos in Gaussian models on statistical manifolds of negative Ricci
curvature}, Int. J. Theor. Phys. \textbf{47}, 2924 (2008).

\bibitem {aliphysica10}S. A. Ali, C. Cafaro, D.-H Kim, and S. Mancini,
\emph{The effect of microscopic correlations on the information geometric
complexity of Gaussian statistical models}, Physica \textbf{A389}, 3117 (2010).

\bibitem {catichaaip07}A. Caticha and C. Cafaro, \emph{From information
geometry to Newtonian dynamics}\textit{, }AIP Conf. Proc. \textbf{954}, 165 (2007).

\bibitem {cafaroejtp08}C. Cafaro and S. A. Ali, \emph{Geometrodynamics of
information on curved statistical manifolds and its applications to chaos},
EJTP \textbf{5}, 139 (2008).

\bibitem {cafarocsf09}C. Cafaro, \emph{Works on an information
geometrodynamical approach to chaos}, Chaos, Solitons \& Fractals \textbf{41},
886 (2009).

\bibitem {zurek94}W. H. Zurek and J. P. Paz, \emph{Decoherence, chaos, and the
second law}, Phys. Rev. Lett. \textbf{72}, 2508 (1994).

\bibitem {zurek95}W. H. Zurek and J. P. Paz, \emph{Quantum chaos: a decoherent
definition}, Physica \textbf{D83}, 300 (1995).

\bibitem {cafaromplb08}C. Cafaro, \emph{Information geometry, inference
methods and chaotic energy levels statistics}, Mod. Phys. Lett. \textbf{B22},
1879 (2008).

\bibitem {cafarophysica08}C. Cafaro and S. A. Ali, \emph{Can chaotic quantum
energy levels statistics be characterized using information geometry and
inference methods?} Physica \textbf{A387}, 6876 (2008).

\bibitem {prosenpre07}T. Prosen and M. Znidaric, \emph{Is the efficiency of
classical simulations of quantum dynamics related to integrability?}, Phys.
Rev. \textbf{E75}, 015202 (2007).

\bibitem {prosenpra07}T. Prosen and I. Pizorn, \emph{Operator space
entanglement entropy in transverse Ising chain}, Phys. Rev. \textbf{A76},
032316 (2007).

\bibitem {cafarops10}C. Cafaro and S. Mancini, \emph{On the complexity of
statistical models admitting correlations}, Physica Scripta \textbf{82},
035007 (2010).

\bibitem {cafaropd11}C. Cafaro and S. Mancini, \emph{Quantifying the
complexity of geodesic paths on curved statistical manifolds through
information geometric entropies and Jacobi fields}, Physica \textbf{D240}, 607 (2011).

\bibitem {kim11}D.-H Kim, S. A. Ali, C. Cafaro, and S. Mancini, \emph{An
information geometric analysis of entangled continuous variable quantum
systems}, Journal of Physics: Conference Series \textbf{306}, 012063 (2011).

\bibitem {kimpla11}D.-H. Kim, S. A. Ali, C. Cafaro\textbf{,} and S.\textbf{
}Mancini, \emph{Information geometric modeling of scattering induced quantum
entanglement}, Phys. Lett. \textbf{A375}, 2868 (2011).

\bibitem {kim12}D.-H. Kim, S.\ A. Ali, C. Cafaro\textbf{, }and S\textbf{.
}Mancini, \emph{Information geometry of quantum entangled wave-packets},
Physica \textbf{A391, }4517 (2012).

\bibitem {cafaroaip12}C. Cafaro, A. Giffin, C. Lupo, and S. Mancini,
\emph{Insights into the softening of chaotic statistical models by quantum
considerations}, AIP Conf. Proc. \textbf{1443}, 366 (2012).

\bibitem {aliaip12}Ali, S. A.; Cafaro, C.; Giffin, A.; Lupo, C.; Mancini, S.
On a differential geometric viewpoint of Jaynes' MaxEnt method and its quantum
extension. \textit{AIP Conf. Proc}. \textbf{2012}, 1443, 120.

\bibitem {giffinaip13}A. Giffin, S. A. Ali, and C. Cafaro,\emph{ Local
softening of chaotic statistical models with quantum consideration}, AIP Conf.
Proc. \textbf{1553}, 238 (2013).

\bibitem {cafaroosid12}C. Cafaro, A. Giffin, C. Lupo, and S. Mancini,
\emph{Softening the complexity of entropic motion on curved statistical
manifolds}, Open Systems \& Information Dynamics \textbf{19}, 1250001 (2012).

\bibitem {giffinentropy13}A. Giffin, S. A. Ali, and C. Cafaro, \emph{Local
softening of information geometric indicators of chaos in statistical modeling
in the presence of quantum-like considerations}, Entropy \textbf{15}, 4622 (2013).

\bibitem {felice14}D. Felice, C. Cafaro, and S. Mancini, Information geometric
complexity of a trivariate Gaussian statistical model, Entropy \textbf{16},
2944 (2014).

\bibitem {sadoc06}J. F. Sadoc and R. Mosseri, \emph{Geometrical Frustration},
Cambridge University Press (2006).

\bibitem {greven}A. Greven, G. Keller, and G. Warnecke, \emph{Entropy},
Princeton University Press (2003).

\bibitem {peres}A. Peres, \emph{Quantum Theory: Concepts and Methods}, Kluwer
Academic Publishers (1995).

\bibitem {peng}L. Peng, H. Sun, and G. Xu, \emph{Information geometric
characterization of the complexity of fractional Brownian motion}, J. Math.
Phys. \textbf{53}, 123305 (2012).

\bibitem {peng2}L. Peng, H. Sun, D. Sun, and J. Yi, \emph{The geometric
structures and instability of entropic dynamical models}, Adv. Math.
\textbf{227}, 459 (2011).

\bibitem {r1}O. Semarak and P. Sukova, \emph{Free motion around black holes
with discs or rings: between integrability and chaos-I}, Monthly Notices of
the Royal Astronomical Society \textbf{404}, 545 (2010).

\bibitem {r2}C. Li, H. Sun, and S. Zhang, \emph{Characterization of the
complexity of an ED model via information geometry}, Eur. Phys. J. Plus
\textbf{128}, 70 (2013).

\bibitem {r3}L. Cao, D. Li, E. Zhang, and H. Sun, \emph{A statistical
cohomogeneity one metric on the upper plane with constant negative curvature},
Adv. Math. Phys. \textbf{2014}, 832683 (2014).

\bibitem {FMP}D. Felice, S. Mancini, and M. Pettini, \emph{Quantifying
networks complexity from information geometry viewpoint}, J. Math. Phys.
\textbf{55}, 043505 (2014).

\bibitem {r5}S. M. Abtahi, S. H.\ Sadati, and H. Salarieh, \emph{Ricci-based
chaos analysis for roto-translatory motion of a Kelvin-type gyrostat
satellite}, Journal of Multi-Body Dynamics \textbf{228}, 34 (2014).

\bibitem {r6}J. Mikes and E. Stepanova, \emph{A five-dimensional Riemannian
manifold with an irreducible SO(3)-structure as a model of abstract
statistical manifold}, Annals of Global Analysis and Geometry \textbf{45}, 111 (2014).

\bibitem {r7}S. Weis, \emph{Continuity of the maximum-entropy inference},
Commun. Math. Phys. \textbf{330}, 1263 (2014).

\bibitem {r8}C. Li, L. Peng, and H. Sun, \emph{Entropic dynamical models with
unstable Jacobi fields}, Rom. Journ. Phys. \textbf{60}, 1249 (2015).

\bibitem {r9}M. Itoh and H. Satoh, \emph{Geometry of Fisher information metric
and the barycenter map}, Entropy \textbf{17}, 1814 (2015).

\bibitem {r10}R. Franzosi, D. Felice, S. Mancini, and M. Pettini, \emph{A
geometric entropy detecting the Erd\"{o}s-R\'{e}nyi phase transition}, Eur.
Phys. Lett. \textbf{111}, 20001 (2015).

\bibitem {r11}A. C. R. Martins, \emph{Opinion particles: Classical physics and
opinion dynamics}, Phys. Lett. \textbf{A379}, 89 (2015).

\bibitem {r12}S. A. Muhammad, E. Zhang, and H. Sun, Jacobi fields on the
manifold of Freund, Italian Journal of Pure and Applied Mathematics
\textbf{34}, 181 (2015).

\bibitem {r13}D. Felice and S. Mancini, \emph{Gaussian network's dynamics
reflected into geometric entropy}, Entropy \textbf{17}, 5660 (2015).

\bibitem {r14}C. Wen-Haw, \emph{A review of geometric mean of positive
definite matrices}, British Journal of Mathematics and Computer
Science\textbf{ 5}, 1 (2015).

\bibitem {r15}S. Weis, A. Knauf, N. Ay, and M.-J. Zhao, \emph{Maximizing the
divergence from a hierarchical model of quantum states}, Open Syst. Inf. Dyn.
\textbf{22}, 1550006 (2015).

\bibitem {r16}S. Weis, \emph{Maximum-entropy inference and inverse continuity
of the numerical range}, Reports on Mathematical Physics\textit{ }\textbf{77},
251 (2016).

\bibitem {r17}D. S. Shalymov and A. L. Fradkov, \emph{Dynamics of
non-stationary processes that follow the maximum of the R\'{e}nyi entropy
principle}, Proc. R. Soc. \textbf{A472}, 20150324 (2016).

\bibitem {r19}G. Henry and D. Rodriguez, \emph{On the instability of two
entropic dynamical models}, Chaos,\ Solitons \& Fractals \textbf{91}, 604 (2016).

\bibitem {r18}I. S. Gomez and M. Portesi, \emph{Ergodic statistical models:
entropic dynamics and chaos}, AIP Conf. Proc. \textbf{1853}, 100001 (2017).

\bibitem {r18a}I. S. Gomez, \emph{Notions of the ergodic hierarchy for curved
statistical manifolds}, Physica \textbf{A484}, 117 (2017).

\bibitem {steven}S. Gassner and C. Cafaro, \emph{Information geometric
complexity of entropic motion on curved statistical manifolds under different
metrizations of probability spaces}, Int. J. Geometric Methods in Modern
Physics \textbf{16}, 1950082 (2019).

\bibitem {gomez20}I. S. Gomez, M. Portesi, and E. P. Borges,
\emph{Universality classes for the Fisher metric derived from relative group
entropy}, Physica \textbf{A547}, 123827 (2020).

\bibitem {summers20}R. L.\ Summers, \emph{Experiences in the Biocontinuum: A
New Foundation for Living Systems}, Cambridge Scholar Publishing (2020).

\bibitem {desh21}S. Deshmukh, A. Ishan, S. B. Al-Shaik, and C. \"{O}zg\"{u}r,
\emph{A note on Killing calculus on Riemannian manifolds}, Mathematics
\textbf{9}, 307 (2021).
\end{thebibliography}
\end{document}